\documentclass[%
 reprint,
 amsmath,amssymb,
 aps,
]{revtex4-2}
\usepackage{mathptmx}
\usepackage{etoolbox}
\usepackage{xcolor}
\usepackage{float}
\usepackage{hyperref}
\usepackage{graphicx}% Include figure files
\usepackage{dcolumn}% Align table columns on decimal point
\usepackage{bm}% bold math
\begin{document}

%\preprint{APS/123-QED}

\title{Electron–Phonon Coupling and Charge-Density-Wave Instabilities in W$_2$N and Halogen-Functionalized W$_2$N Monolayers}% Force line breaks with \\

\author{Jakkapat Seeyangnok$^{1}$}
 \email{jakkapatjtp@gmail.com} 
 %\altaffiliation[Also at ]{Department of Physics, Faculty of Science, Chulalongkorn University, Bangkok, Thailand.}%Lines break automatically or can be forced with \\
\author{Udomsilp Pinsook$^{1}$}%
 \email{Udomsilp.P@Chula.ac.th}
%\affiliation{Department of Physics, Faculty of Science, Chulalongkorn University, Bangkok, Thailand.}%

%\author{Graeme J Ackland$^{2}$}
%\email{gjackland@ed.ac.uk} 
\affiliation{$^{1}$Department of Physics, Faculty of Science, Chulalongkorn University, Bangkok, Thailand.}
%^{2}$Centre for Science at Extreme Conditions, School of Physics and Astronomy, University of Edinburgh, Edinburgh, United Kingdom}%


\date{\today}% It is always \today, today,

\begin{abstract}
The interplay between charge-density-wave (CDW) order and superconductivity is a central problem in condensed-matter physics because both phenomena often originate from the same electron--phonon coupling (EPC) mechanism. Here, we investigate the structural, electronic, vibrational, and superconducting properties of monolayer W$_2$N and halogen-functionalized W$_2$N ($\mathrm{W_2NF_2}$ and $\mathrm{W_2NCl_2}$) using first-principles calculations. Pristine W$_2$N exhibits pronounced phonon instabilities near the M and K points driven by exceptionally strong EPC associated with softened low-frequency phonons. The coincidence between phonon softening and enhanced phonon linewidths identifies the instability as EPC-driven and indicative of a CDW tendency. Inclusion of van der Waals interactions stabilizes the lattice and yields strong-coupling superconductivity with $\lambda=1.00$ and $T_c=13.2$ K, while fluorination further weakens the soft-phonon anomaly, resulting in a moderate-coupling superconductor with $\lambda=0.67$ and $T_c=5.3$ K. In contrast, W$_2$NCl$_2$ exhibits a re-emergence of CDW-related phonon softening that can be continuously suppressed by compressive strain or electron doping. Under $-3\%$ compressive strain, the EPC constant decreases from $\lambda=1.35$ to $\lambda=0.71$, giving rise to superconductivity with $T_c=5.8$ K. Across the entire W$_2$N family, the low-energy physics is governed by softened ZA phonons near the M point, establishing a unified framework in which CDW order and superconductivity emerge as competing manifestations of the same soft-phonon-driven EPC mechanism.
\end{abstract}

%Use showkeys class option if keyword
\maketitle

%\tableofcontents
% =========================
\section{Introduction}
% =========================
Two-dimensional (2D) materials provide an ideal platform for exploring emergent quantum phenomena arising from reduced dimensionality and enhanced many-body interactions. Since the discovery of graphene, atomically thin materials have been shown to host a rich variety of collective electronic states, including superconductivity, magnetism, topological phases, and charge-density waves (CDWs)~\cite{balandin2021charge}. Among these phenomena, the interplay between superconductivity and CDW order has attracted sustained interest because both phases emerge from electronic states near the Fermi level and are often closely linked through electron--phonon coupling (EPC)~\cite{wang2023interplay,ali2025interplay}. Understanding how EPC drives, stabilizes, or suppresses these competing collective states remains a central challenge in condensed-matter physics.

In recent years, chemical functionalization has emerged as an effective approach for engineering the electronic and vibrational properties of 2D materials~\cite{bekaert2019hydrogen,seeyangnok2026enhanced,liu2024three,seeyangnok2026stability,seeyangnok2026tunable}. By modifying the local bonding environment and lattice dynamics, surface functionalization can strongly influence phonon spectra and EPC, thereby tuning superconductivity and structural instabilities. Hydrogenation, in particular, has been shown to significantly enhance EPC and induce phonon-mediated superconductivity in a wide range of low-dimensional materials~\cite{wang2023hydrogenation,seeyangnok2025high,meng2025mbenes,seeyangnok2025ab,seeyangnok2026theoretical,xue2024realization,seeyangnok2025hydrogenation,han2023high,seeyangnok2025phase}. Prominent examples include Janus transition-metal chalcogenide hydrides such as MoSH, MoSeH, WSH, and WSeH, as well as related Janus $MX$H monolayers ($M=\mathrm{Ti,Zr,Hf}$ and $X=\mathrm{S,Se,Te}$), where strong EPC gives rise to superconductivity with a wide range of transition temperatures~\cite{lu2017janus,liu2022two,sui2025two,seeyangnok2024superconductivity,seeyangnok2024superconductivitywseh,qiao2024prediction,li2024machine,seeyangnok2025competition}.

Strong EPC, however, can also drive lattice instabilities and the formation of CDW order. A CDW state is characterized by a periodic modulation of the electronic charge density accompanied by a symmetry-breaking lattice distortion~\cite{gruner2018density}. While CDWs were originally interpreted within the Peierls picture of Fermi-surface nesting~\cite{peierls1955quantum}, it is now widely recognized that CDW formation in most two-dimensional materials is primarily governed by momentum-dependent EPC and selective phonon softening~\cite{johannes2008fermi,zhu2017misconceptions}. This mechanism underlies the CDW behavior observed in numerous layered compounds, including NbS$_2$, NbSe$_2$, TaSe$_2$, TaS$_2$, 2H-Pd$_x$TaSe$_2$, Mo$X$H ($X$=S,Se), and the recently predicted monolayer Mo$_2$NF$_2$~\cite{hwang2024charge,weber2011extended,calandra2011charge,ugeda2016characterization,xi2015strongly,lian2023interplay,bhoi2016interplay,seeyangnok2026competition,seeyangnok2026moxhcdw}. In many of these systems, CDW order coexists or competes with superconductivity, highlighting the dual role of EPC in stabilizing distinct collective phases.

Among transition-metal nitrides, monolayer W$_2$N has recently emerged as a promising superconducting material due to its metallic electronic structure and exceptionally strong EPC. However, the presence of pronounced phonon softening indicates that the system lies in close proximity to a lattice instability. Moreover, the influence of surface functionalization and external tuning parameters on the competition between CDW order and superconductivity in W$_2$N-based systems remains largely unexplored. In particular, halogen functionalization offers an attractive route for modifying the lattice dynamics and EPC strength, potentially enabling systematic control over collective quantum phases.

In this work, we investigate the structural, electronic, vibrational, and superconducting properties of W$_2$N, vdW-corrected W$_2$N, W$_2$NF$_2$, and W$_2$NCl$_2$ using first-principles calculations. We demonstrate that all four systems can be understood within a unified framework governed by soft-phonon-driven EPC. Pristine W$_2$N exhibits a strong EPC-induced CDW instability, while vdW interactions stabilize the lattice and yield strong-coupling superconductivity. Fluorination further suppresses the soft-phonon anomaly, producing a moderate-coupling superconducting state. In contrast, W$_2$NCl$_2$ lies near a CDW instability, which can be continuously tuned and suppressed through external strain and carrier doping. Our results reveal how lattice stability, CDW order, and superconductivity evolve across the W$_2$N family and establish these monolayers as a versatile platform for studying EPC-driven quantum phases in two-dimensional materials.

% =========================
\section{Computational Methods}
% =========================
First-principles calculations were carried out within the framework of density functional theory (DFT) using the \textsc{Quantum ESPRESSO} package~\cite{giannozzi2009quantum}. Exchange--correlation interactions were described within the generalized gradient approximation (GGA) employing the Perdew--Burke--Ernzerhof (PBE) functional~\cite{perdew1996generalized}, together with optimized norm-conserving Vanderbilt (ONCV) pseudopotentials~\cite{schlipf2015optimization}. Plane-wave kinetic-energy and charge-density cutoffs of 80 and 320 Ry were adopted, respectively. To eliminate spurious interactions between periodically repeated layers, a vacuum spacing of 20~\AA\ was introduced along the out-of-plane direction. Structural relaxations were performed until the residual Hellmann--Feynman forces on all atoms were smaller than $10^{-5}$~eV/\AA. Brillouin-zone integrations were carried out using a $24\times24\times1$ Monkhorst--Pack $\mathbf{k}$-point mesh~\cite{monkhorst1976special} together with a Methfessel--Paxton smearing width of 0.02 Ry~\cite{methfessel1989high}.

Phonon spectra and electron--phonon coupling (EPC) properties were evaluated within density functional perturbation theory (DFPT)~\cite{baroni2001phonons}. Dynamical matrices were calculated on $12\times12\times1$ $\mathbf{q}$-point grids for the primitive and supercell structures, respectively. The interaction between electrons and phonons gives rise to a finite phonon linewidth,

\begin{equation}
\gamma_{\mathbf{q}\nu}
=
2\pi\omega_{\mathbf{q}\nu}
\sum_{nm}
\sum_{\mathbf{k}}
\left|
g_{\mathbf{k}+\mathbf{q},\mathbf{k}}^{\mathbf{q}\nu,mn}
\right|^2
\delta(\epsilon_{\mathbf{k}+\mathbf{q},m}-\epsilon_F)
\delta(\epsilon_{\mathbf{k},n}-\epsilon_F),
\label{gammaphononlinewidths}
\end{equation}

where $g_{\mathbf{k}+\mathbf{q},\mathbf{k}}^{\mathbf{q}\nu,mn}$ is the electron--phonon matrix element and $\epsilon_F$ denotes the Fermi energy. The mode-resolved EPC parameter is then given by

\begin{equation}
\lambda_{\mathbf{q}\nu}
=
\frac{\gamma_{\mathbf{q}\nu}}
{\pi N(\epsilon_F)\omega_{\mathbf{q}\nu}^{2}},
\end{equation}

where $N(\epsilon_F)$ represents the electronic density of states at the Fermi level and $\omega_{\mathbf{q}\nu}$ is the phonon frequency of branch $\nu$ at wave vector $\mathbf{q}$.

The superconducting transition temperature $T_c$ was estimated from the isotropic Eliashberg spectral function $\alpha^2F(\omega)$ using the Allen--Dynes modified McMillan equation~\cite{allen1975transition,pinsook2024analytic},

\begin{equation}
T_c
=
\frac{f_1 f_2 \omega_{\ln}}{1.2}
\exp
\left[
-\frac{1.04(1+\lambda)}
{\lambda-\mu^{\ast}(1+0.62\lambda)}
\right],
\end{equation}

where $\lambda$ is the total EPC constant, $\omega_{\ln}$ is the logarithmic average phonon frequency, and $\mu^{\ast}$ is the Coulomb pseudopotential. The EPC constant and logarithmic phonon frequency were obtained from the Eliashberg spectral function through

\begin{equation}
\lambda
=
2\int_{0}^{\infty}
\frac{\alpha^2F(\omega)}
{\omega}
\, d\omega,
\end{equation}

and

\begin{equation}
\omega_{\ln}
=
\exp
\left[
\frac{2}{\lambda}
\int_{0}^{\infty}
\frac{\alpha^2F(\omega)}
{\omega}
\ln(\omega)
\, d\omega
\right].
\end{equation}

Unless otherwise specified, a Coulomb pseudopotential of $\mu^{\ast}=0.10$ was used throughout this work.

% =========================
\section{Results and Discussion}
% =========================

% =========================
\subsection{Charge-Density-Wave Instability in Monolayer W$_2$N}
% =========================
    \begin{figure}[h!]
        \centering
        \includegraphics[width=8.5cm]{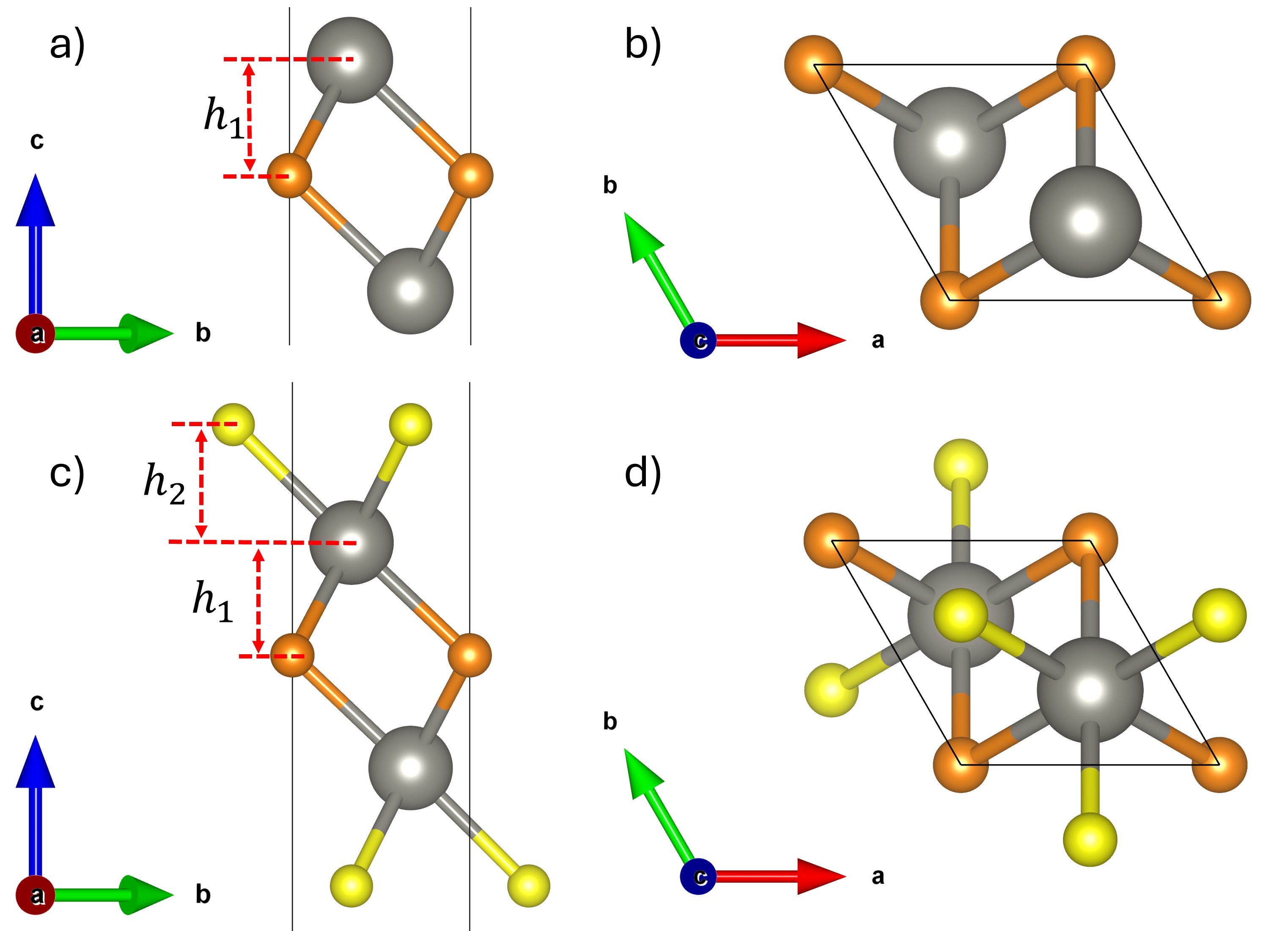}
        \caption{Crystal structures of pristine W$_2$N and functionalized W$_2$NT$_2$ ($T=$ F, Cl) monolayers belonging to the trigonal $P\bar{3}m1$ (No.~164) space group. (a) Side and (b) top views of pristine W$_2$N. The structural parameter $h$ denotes the vertical separation between the N and W atomic planes. (c) Side and (d) top views of the functionalized W$_2$NT$_2$ monolayers. The parameters $h_1$ and $h_2$ represent the vertical distances between the N layer and the W layer, and between the W layer and the terminating halogen layer, respectively. Gray, orange, and yellow spheres correspond to W, N, and halogen atoms, respectively. The primitive unit cell is outlined by solid black lines.}
        \label{fig:structure}
    \end{figure}

    \begin{figure*}[ht]
        \centering
        \includegraphics[width=18cm]{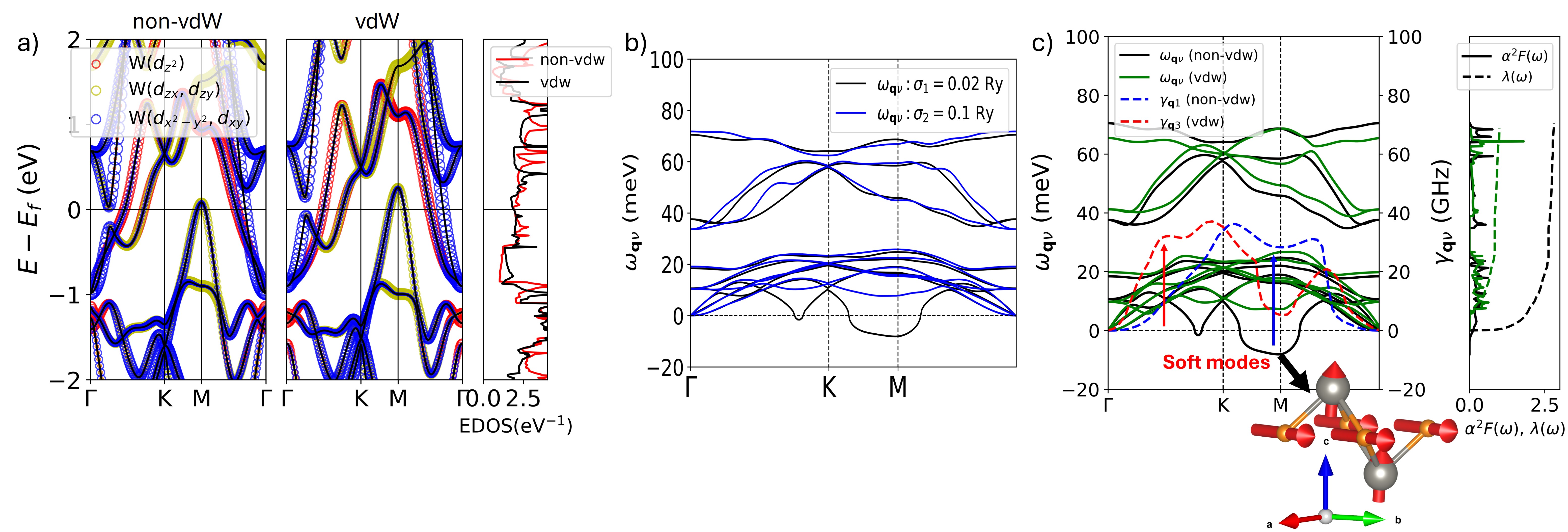}
        \caption{(a) Orbital-projected electronic band structures, Fermi surfaces, and electronic density of states (EDOS) of monolayer W$_2$N calculated without and with van der Waals (vdW) corrections. The states near the Fermi level are dominated by W-$d$ orbitals, confirming metallic behavior. The Fermi-surface topology is preserved upon inclusion of vdW interactions, although the hole pocket around the M point becomes larger. (b) Phonon dispersions of W$_2$N calculated with electronic smearing parameters $\sigma=0.02$ and $0.10$ Ry, showing stabilization of the soft phonon modes at larger smearing. (c) Phonon dispersions and phonon linewidths $\gamma_{\mathbf{q}\nu}$ for the non-vdW and vdW structures. The non-vdW phase exhibits imaginary phonon modes near K and M, whereas vdW interactions stabilize these instabilities into low-frequency soft modes. The maxima in $\gamma_{\mathbf{q}\nu}$ coincide with the soft-phonon wave vectors, indicating strong electron--phonon coupling (EPC). The right panel shows the Eliashberg spectral function $\alpha^2F(\omega)$ and cumulative EPC constant $\lambda(\omega)$ of the vdW-stabilized phase. The inset depicts the atomic displacement pattern of the soft mode.}
        \label{fig:w2n-electronic-phonon}
    \end{figure*}

Pristine W$_2$N and halogen-functionalized W$_2$NT$_2$ ($T=$ F, Cl) monolayers crystallize in the trigonal $P\bar{3}m1$ (No.~164) space group, as illustrated in Fig.~\ref{fig:structure}. The primitive cell contains two W atoms and one N atom for W$_2$N, while functionalization introduces two additional halogen atoms located on opposite sides of the monolayer. The optimized structures belong to the trigonal $P\bar{3}m1$ (No.~164) space group. In the primitive unit cell, the N atom occupies the Wyckoff position $(0,0,0)$, while the two W atoms are located at $(2/3,1/3,\pm z_{\mathrm{W}})$. Upon halogen functionalization, two additional $T$ atoms ($T=$ F, Cl) are adsorbed on opposite sides of the monolayer at $(1/3,2/3,\pm z_T)$, resulting in the W$_2$NT$_2$ structure while maintaining the overall crystal symmetry.

    \begin{table}[h!]
    \centering
    \caption{Optimized structural parameters of pristine W$_2$N and functionalized W$_2$NT$_2$ ($T=$ F, Cl) monolayers. The lattice constant $a$, W--N vertical separation $h$ (or $h_1$), and W--$T$ vertical separation $h_2$ are given in \AA.}
    \label{tab:lattice}
    \begin{tabular}{lcccc}
    \hline\hline
    System & Method & $a$ (\AA) & $h_1$ (\AA) & $h_2$ (\AA) \\
    \hline
    W$_2$N      & non-vdW & 2.77 & 1.46 & -- \\
    W$_2$N      & vdW     & 2.73 & 1.50 & -- \\
    W$_2$NF$_2$ & vdW     & 2.73 & 1.50 & 1.57 \\
    W$_2$NCl$_2$& vdW     & 2.90 & 1.47 & 1.86 \\
    \hline\hline
    \end{tabular}
    \end{table}

The optimized lattice parameters obtained from density-functional theory are summarized in Table~\ref{tab:lattice}. For pristine W$_2$N, calculations without van der Waals (vdW) corrections yield an in-plane lattice constant of $a=2.77$~\AA\ and a layer thickness of $h_1 =1.46$~\AA. These values are essentially unchanged when increasing the electronic smearing from $\sigma=0.02$ to $0.10$~Ry, indicating that the equilibrium structure is insensitive to the smearing parameter. Inclusion of vdW corrections slightly contracts the lattice to $a=2.73$~\AA\ and increases the layer thickness to $h_1 =1.50$~\AA. Upon fluorination, the lattice constant remains nearly unchanged at $a=2.73$~\AA, while the out-of-plane geometry becomes asymmetric with $h_1=1.50$~\AA\ and $h_2=1.57$~\AA. In contrast, chlorination leads to a noticeable lattice expansion to $a=2.90$~\AA, accompanied by a significantly larger halogen displacement from the W layer ($h_2=1.86$~\AA), reflecting the larger ionic radius of Cl compared with F. The W--N distance remains relatively unchanged ($h_1=1.47$~\AA), indicating that halogen functionalization primarily affects the outer atomic layers while preserving the metallic W$_2$N framework.

Figure~\ref{fig:w2n-electronic-phonon}(a) shows the orbital-projected electronic band structures of monolayer W$_2$N obtained with and without vdW corrections. In both cases, multiple bands cross the Fermi level, demonstrating a robust metallic character. The electronic states near $E_F$ are dominated by W-$d$ orbitals, primarily originating from the $d_{z^2}$, $d_{xz/yz}$, and $d_{x^2-y^2}/d_{xy}$ manifolds. The corresponding Fermi surfaces reveal nearly identical topologies for the two calculations, indicating that vdW interactions do not qualitatively alter the low-energy electronic structure. However, the vdW-relaxed structure exhibits a noticeably larger hole pocket around the M point, suggesting a modification of the electronic states involved in electron--phonon scattering.

To further examine the chemical bonding, Bader charge and integrated crystal orbital Hamilton population (ICOHP) analyses were performed. For the non-vdW structure, each W atom donates approximately $0.77e$, resulting in a charge gain of $1.55e$ on the N atom. Upon inclusion of vdW interactions, the charge transfer slightly decreases to $0.71e$ per W atom and $1.42e$ on N, indicating a reduced ionic character. Consistent with this trend, the W--N bond remains significantly stronger than the W--W interaction, with ICOHP values of $-4.60$ and $-0.36$ eV for the non-vdW structure, respectively. For the vdW-relaxed structure, the corresponding values become $-4.51$ and $-0.27$ eV, indicating slightly weaker W--N and W--W bonding interactions. These results suggest that the vdW-induced stabilization discussed below is not primarily driven by stronger chemical bonding. Instead, it originates from subtle vdW-induced modifications of the equilibrium crystal structure, which alter the electronic states near the Fermi level and consequently weaken the electron--phonon-driven lattice instability.

The phonon dispersions of W$_2$N are presented in Fig.~\ref{fig:w2n-electronic-phonon}(b,c). For the non-vdW structure, pronounced imaginary phonon frequencies appear near the M and K points. The instability near M originates from the lowest acoustic branch ($\nu=1$), whereas an additional instability near K is associated with the $\nu=3$ branch. Such soft phonon modes are characteristic signatures of a charge-density-wave (CDW) instability. The origin of these instabilities can be identified from the corresponding phonon linewidths $\gamma_{\mathbf{q}\nu}$ shown in Fig.~\ref{fig:w2n-electronic-phonon}(c). The pronounced phonon linewidths $\gamma_{\mathbf{q}\nu}$ occur in the vicinity of the wave vectors where the phonon frequencies soften, indicating that the instability is driven by strong electron--phonon coupling rather than by purely structural effects. This behavior is analogous to EPC-driven CDW systems reported previously~\cite{seeyangnok2026competition,seeyangnok2026moxhcdw,ku2023ab,seeyangnok2026tunable,lian2023interplay}.

The EPC origin of the instability is further supported by the dependence on electronic smearing. As shown in Fig.~\ref{fig:w2n-electronic-phonon}(b), increasing the smearing parameter from $\sigma=0.02$ to $0.10$ Ry completely removes the imaginary frequencies and stabilizes the lattice. Physically, larger electronic broadening weakens the sharp electronic scattering processes responsible for EPC enhancement and Fermi-surface nesting, thereby suppressing the phonon instability, in agreement with previous studies of CDW materials~\cite{ku2023ab}. To explore the possibility of a static CDW ground state, we performed structural relaxations using several commensurate supercells, including $2\times2\times1$, $3\times3\times1$, and $4\times4\times1$, initialized with symmetry-breaking distortions. However, no lower-energy distorted structure was identified, and the relaxed configurations remained close to the high-symmetry phase. This result suggests that the EPC-driven phonon instability may be associated with a shallow energy landscape or dynamic CDW fluctuations rather than a strongly stabilized commensurate CDW state.

Interestingly, inclusion of vdW interactions also stabilizes the lattice while preserving strong EPC signatures. The imaginary modes evolve into low-frequency positive-energy soft phonons located at essentially the same wave vectors as the original instabilities. This indicates that the underlying EPC remains strong, but the lattice is sufficiently renormalized to avoid a static CDW distortion. The vdW-stabilized phase therefore represents a regime in which the EPC is large enough to enhance superconductivity while remaining below the threshold required for CDW formation.

The resulting superconducting properties are characterized by an EPC constant $\lambda=1.00$, logarithmic phonon frequency $\omega_{\log}=167.91$ K, and second-moment frequency $\omega_2=288.62$ K, yielding a superconducting critical temperature of $T_c=13.2$ K for $\mu^*=0.10$. The large EPC originates primarily from the strongly softened low-frequency phonons, consistent with the general understanding that Kohn-like phonon anomalies can significantly enhance superconductivity~\cite{jiang2023possible}. In contrast, the unstable non-vdW structure exhibits a much larger EPC constant of $\lambda=2.81$, indicating that the system lies beyond the stability limit and tends toward CDW formation. These results suggest that the CDW instability and superconductivity originate from the same underlying EPC mechanism. Rather than eliminating superconductivity, the CDW instability acts as a competing lattice-relaxation channel that reduces the excessive EPC strength and stabilizes the crystal. In W$_2$N, vdW interactions effectively tune the system from an EPC-driven CDW instability toward a stable superconducting phase without requiring external perturbations such as strain, pressure, or electrostatic gating.

    \begin{figure}[h!]
        \centering
        \includegraphics[width=8.6cm]{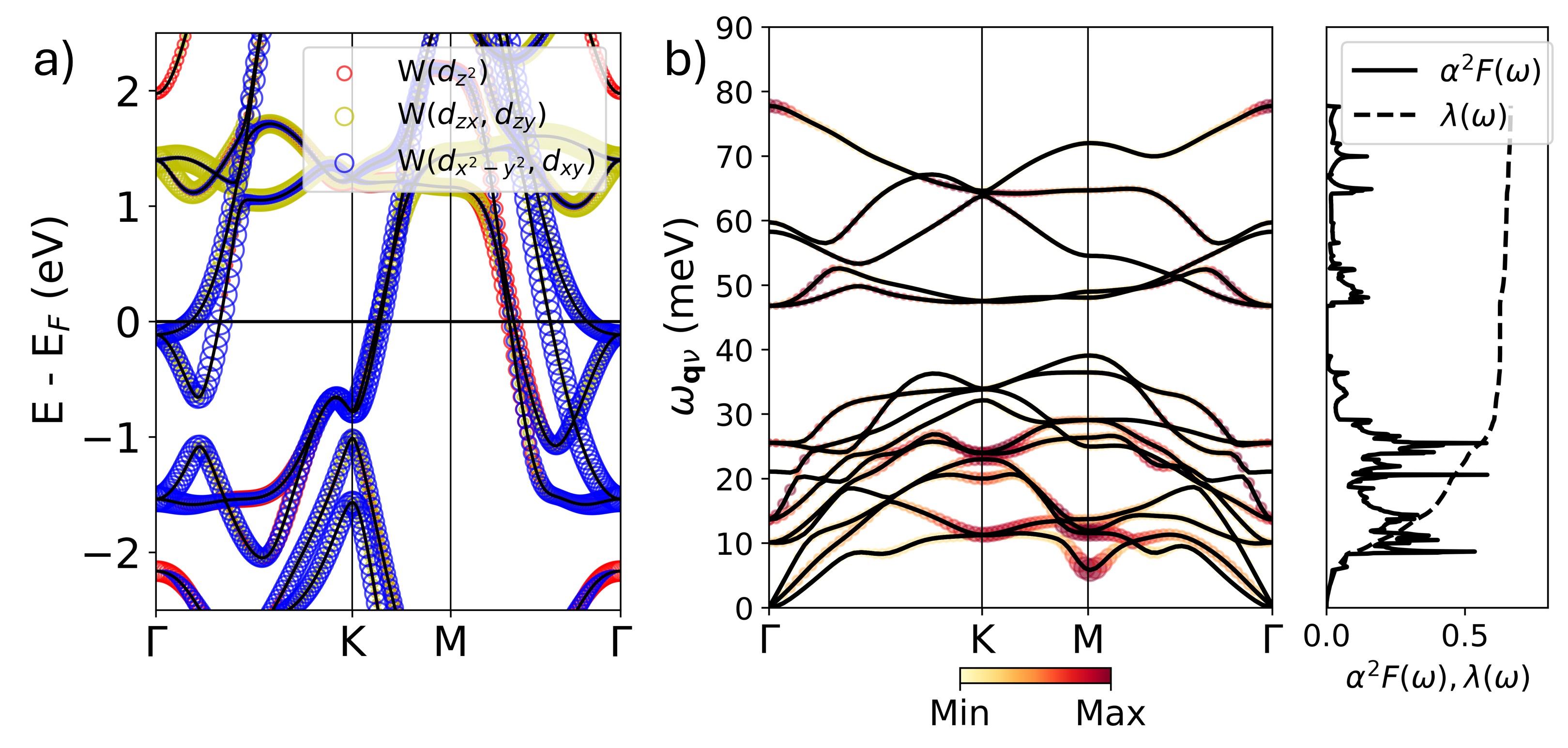}
        \caption{(a) Orbital-projected electronic band structure of monolayer W$_2$NF$_2$. The size of the circles is proportional to the contribution of the corresponding W-$d$ orbitals. Multiple bands cross the Fermi level, confirming the metallic nature of the system. The low-energy electronic states are dominated by W-$d{z^2}$, $d{xz/yz}$, and $d_{x^2-y^2}/d_{xy}$ orbitals. (b) EPC-weighted phonon dispersion of W$_2$NF$_2$, where the color scale represents the mode-resolved electron--phonon coupling strength $\lambda{\mathbf{q}\nu}$. A pronounced low-frequency soft phonon is observed around the M point, primarily associated with the ZA ($\nu=1$) branch. The right panel shows the Eliashberg spectral function $\alpha^2F(\omega)$ and cumulative EPC constant $\lambda(\omega)$.}
        \label{fig:w2nf2}
    \end{figure}
% =========================
\subsection{Superconductivity and Persistent Soft-Phonon Signatures in W$_2$NF$_2$}
% =========================
The electronic structure of W$_2$NF$_2$ is shown in Fig.~\ref{fig:w2nf2}(a). Similar to pristine W$_2$N, the states near the Fermi level are predominantly derived from W-$d$ orbitals, and several bands cross $E_F$, indicating that fluorination preserves the metallic character of the system. The orbital-projected band structure reveals substantial contributions from the $d_{z^2}$, $d_{xz/yz}$, and $d_{x^2-y^2}/d_{xy}$ orbitals, which dominate the low-energy electronic excitations and participate in the electron--phonon interaction.

The EPC-weighted phonon dispersion is presented in Fig.~\ref{fig:w2nf2}(b). Similar to vdW-W$_2$N, W$_2$NF$_2$ exhibits a low-frequency soft phonon mode near the M point, indicating the persistence of strong electron--phonon interactions. However, the softening is less pronounced than in vdW-W$_2$N, consistent with a reduced EPC strength. The dominant contribution to the EPC originates from the ZA ($\nu=1$) branch near the M point, where the softened phonon mode carries a substantial fraction of the total coupling.

The Eliashberg spectral function $\alpha^2F(\omega)$ and integrated EPC constant $\lambda(\omega)$ yield an EPC constant of $\lambda=0.67$, a logarithmic phonon frequency $\omega_{\log}=169.28$ K, and a second-moment frequency $\omega_2=251.97$ K. Using the Allen--Dynes modified McMillan equation with $\mu^*=0.10$, we obtain a superconducting critical temperature of $T_c=5.3$ K. Compared with vdW-W$_2$N ($\lambda=1.00$, $T_c=13.2$ K), fluorination weakens the soft-mode-driven EPC and consequently lowers $T_c$. Nevertheless, W$_2$NF$_2$ remains a moderate-coupling phonon-mediated superconductor, demonstrating that fluorination stabilizes the lattice while retaining the essential EPC mechanism responsible for superconductivity.

% =========================
\subsection{Tunable Suppression of the CDW Instability and Emergence of Superconductivity in W$_2$NCl$_2$}
% =========================
    \begin{figure*}[ht]
        \centering
        \includegraphics[width=18cm]{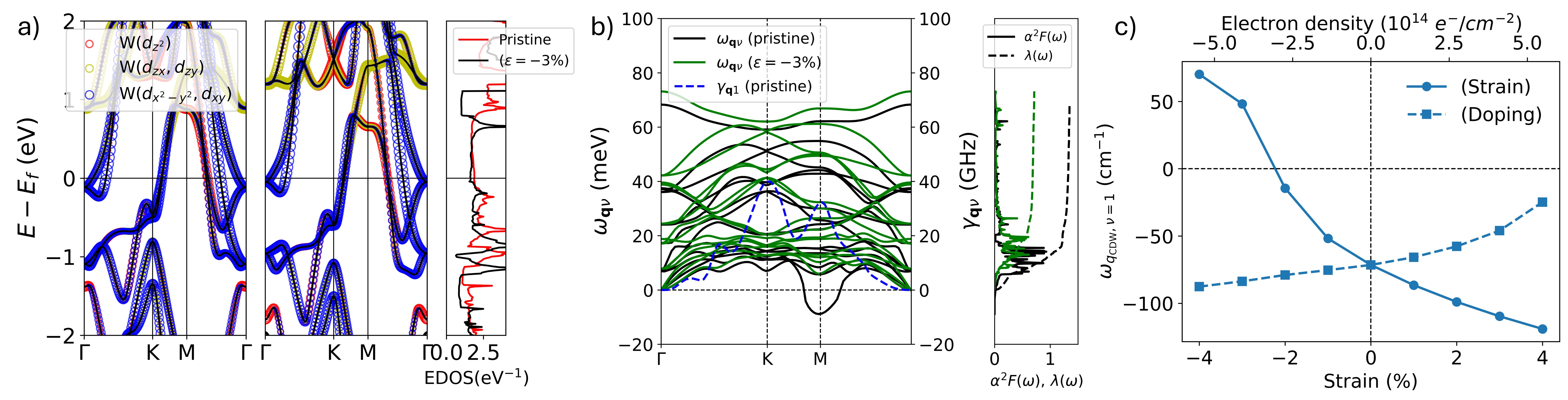}
        \caption{(a) Orbital-projected electronic band structures and EDOS of pristine W$_2$NCl$_2$ and W$_2$NCl$2$ under $-3\%$ compressive strain. Circle sizes are proportional to the corresponding W-$d$ orbital weights. (b) Phonon dispersions and phonon linewidth $\gamma{\mathbf{q}\nu}$ of pristine and strained W$2$NCl$2$. The right panel shows the Eliashberg spectral function $\alpha^2F(\omega)$ and cumulative EPC constant $\lambda(\omega)$ for the strained phase. (c) Evolution of the soft-mode frequency $\omega{q{\mathrm{CDW}},\nu=1}$ under biaxial strain and electron doping. The soft mode is progressively hardened and becomes dynamically stable beyond the critical strain or carrier concentration.}
        \label{fig:w2ncl2-epc}
    \end{figure*}

Figure~\ref{fig:w2ncl2-epc}(a) presents the orbital-projected electronic band structures and electronic density of states (EDOS) of pristine W$2$NCl$2$ and W$2$NCl$2$ under $-3\%$ compressive strain. In both cases, multiple bands cross the Fermi level, confirming the metallic nature of the system. The electronic states around $E_F$ are predominantly composed of W-$d$ orbitals, particularly the $d{z^2}$, $d{xz}/d{yz}$, and $d{x^2-y^2}/d_{xy}$ states. Notably, the overall band dispersions and EDOS remain nearly unchanged upon applying compressive strain, indicating that the electronic structure is relatively insensitive to moderate lattice compression. Therefore, the strain-induced changes in the low-energy properties are unlikely to originate from substantial modifications of the electronic bands.

The phonon dispersions and electron--phonon coupling (EPC) properties are shown in Fig.~\ref{fig:w2ncl2-epc}(b). For pristine W$_2$NCl$2$, an imaginary phonon mode appears at the M point, signaling a dynamical instability of the high-symmetry structure. The phonon linewidth $\gamma_{\mathbf{q}\nu}$ exhibits a pronounced peak at the same wave vector as the soft phonon mode, demonstrating a direct correlation between the lattice instability and strong electron--phonon coupling. This coincidence indicates that the instability is EPC-driven and is associated with a charge-density-wave (CDW) tendency rather than a purely structural instability. A similar supercell analysis was performed for W$_2$NCl$_2$. Despite the presence of an imaginary phonon mode at the M point, structural relaxations of $2\times2\times1$, $3\times3\times1$, and $4\times4\times1$ supercells did not reveal a lower-energy commensurate CDW phase. The instability therefore appears to reflect a strong EPC-driven tendency toward CDW formation, which can be continuously suppressed by compressive strain or electron doping.

Applying a biaxial compressive strain of $-3\%$ completely removes the imaginary phonon frequency, stabilizing the lattice while preserving a low-frequency soft phonon near the M point. The persistence of this softened mode suggests that the system remains close to a CDW instability, although the EPC strength is significantly weakened. Consequently, the total EPC constant is reduced from $\lambda = 1.35$ in the pristine structure to $\lambda = 0.71$ under compressive strain. The Eliashberg spectral function $\alpha^2F(\omega)$ further reveals that the dominant contribution to the remaining EPC originates from low-frequency phonons associated with the softened ZA branch. Using the Allen--Dynes formalism with $\mu^{\ast}=0.10$, together with $\omega_{\log}=160.27$ K and $\omega_2=226.77$ K, we obtain a superconducting transition temperature of $T_c=5.8$ K. These results indicate that compressive strain suppresses the CDW instability while preserving sufficient electron--phonon interaction to support conventional phonon-mediated superconductivity.

To further quantify the evolution of the instability, Fig.~\ref{fig:w2ncl2-epc}(c) shows the dependence of the soft-mode frequency at the CDW wave vector, $\omega_{q_{\mathrm{CDW}},\nu=1}$, on external tuning parameters. Under compressive strain, the magnitude of the imaginary frequency decreases continuously and eventually changes sign, indicating a transition from a dynamically unstable CDW state to a stable metallic phase. A similar trend is observed under electron doping, where increasing carrier concentration progressively hardens the soft mode and suppresses the phonon instability. The continuous evolution of $\omega_{q_{\mathrm{CDW}},\nu=1}$ under both strain and doping demonstrates that the CDW phase in W$_2$NCl$_2$ is highly tunable and originates from an EPC-driven soft-phonon mechanism. Consequently, both lattice compression and carrier injection provide effective routes for controlling the competition between CDW order and superconductivity in this material.

% =========================
\subsection{Unified Picture of Charge-Density-Wave Instabilities in W$_2$N-Based Monolayers}
% =========================
\begin{table*}[ht]
\caption{Summary of the electron--phonon coupling (EPC), superconducting properties, and charge-density-wave (CDW) tendencies of W$_2$N-based monolayers. Imaginary phonon frequencies indicate dynamical instability toward a CDW phase.}
\label{tab:summary}
\centering
\begin{tabular}{lcccccc}
\hline\hline
System & $\lambda$ & $\omega_{\log}$ (K) & $\omega_2$ (K) & $T_c$ (K) & Dominant soft mode & Ground state \\
\hline
W$_2$N (non-vdW)              & 2.81 & --     & --     & --   & ZA(M), $\nu=1$; $\nu=3$(K) & CDW unstable \\
W$_2$N (vdW)                  & 1.00 & 167.91 & 288.62 & 13.2 & Soft ZA(M) & Strong-coupling SC \\
W$_2$NF$_2$ (vdW)             & 0.67 & 169.28 & 251.97 & 5.3  & Soft ZA(M) & Moderate-coupling SC \\
W$_2$NCl$_2$ (vdW)            & 1.35 & --     & --     & --   & ZA(M) & CDW unstable \\
W$_2$NCl$_2$ (vdW, $-3\%$)    & 0.71 & 160.27 & 226.77 & 5.8  & Soft ZA(M) & Moderate-coupling SC \\
\hline\hline
\end{tabular}
\end{table*}
    
The results presented above, summarized in Table~\ref{tab:summary}, reveal a common physical mechanism governing the interplay between charge-density-wave (CDW) order and superconductivity in W$_2$N-based monolayers. In all systems, the electron--phonon coupling (EPC) is dominated by strongly softened low-frequency phonons, particularly the ZA branch near the M point. These soft phonons enhance the EPC strength and simultaneously promote both superconducting pairing and lattice instabilities, consistent with the general role of Kohn-like phonon anomalies in EPC-driven quantum materials~\cite{jiang2023possible}.

Pristine W$_2$N represents the strong-coupling limit of this behavior. The EPC constant reaches $\lambda=2.81$, driving imaginary phonon modes near the M and K points and signaling a dynamical instability. The resulting CDW phase can be viewed as a self-stabilization mechanism in which the lattice lowers its energy through symmetry breaking and suppresses the excessive EPC responsible for the instability. Thus, the CDW and superconducting tendencies originate from the same underlying interaction.

External perturbations systematically tune this balance. Inclusion of vdW interactions stabilizes the lattice by converting the imaginary phonons into low-frequency positive modes, reducing the EPC constant to $\lambda=1.00$ while maintaining strong-coupling superconductivity with $T_c=13.2$ K. Fluorination further weakens the soft phonon anomaly, yielding a moderate EPC strength of $\lambda=0.67$ and a superconducting transition temperature of $T_c=5.27$ K.

Chlorination shifts the system back toward the instability regime. Pristine W$_2$NCl$_2$ exhibits a re-emergence of the soft phonon at the M point accompanied by enhanced phonon linewidth, indicating an EPC-driven CDW instability. Under $-3\%$ compressive strain, however, the imaginary phonon is completely stabilized while retaining a low-frequency soft mode. Consequently, the EPC constant is reduced from $\lambda=1.35$ to $\lambda=0.71$, with $\omega_{\log}=160.27$ K and $\omega_2=226.77$ K, resulting in a superconducting transition temperature of $T_c=5.8$ K for $\mu^{\ast}=0.10$. The soft-mode frequency can be continuously tuned by both strain and electron doping, demonstrating that the CDW instability remains highly controllable through external perturbations.

Overall, W$_2$N, vdW-W$_2$N, W$_2$NF$_2$, and W$_2$NCl$_2$ form a unified family in which the strength of soft-phonon-driven EPC determines whether the system exhibits a CDW instability, a strong-coupling superconducting state, or a moderate-coupling superconducting phase. The CDW and superconducting states should therefore be regarded as two manifestations of the same EPC-driven physics, with external tuning parameters controlling the crossover between them.

% =========================
\section{Conclusions}
% =========================
In summary, we have systematically investigated the interplay between charge-density-wave (CDW) instabilities, electron--phonon coupling (EPC), and superconductivity in W$_2$N-based monolayers using first-principles calculations. We find that the low-energy physics of this family is governed by strongly softened phonon modes, particularly the ZA branch near the M point, which provides the dominant contribution to the EPC and controls the emergence of both CDW order and superconductivity.

Pristine W$_2$N exhibits strong EPC-driven phonon instabilities, indicating a tendency toward CDW formation. The resulting CDW state can be interpreted as a self-stabilization mechanism that removes the excessive EPC through spontaneous translational-symmetry breaking. Inclusion of vdW interactions stabilizes the lattice while retaining substantial EPC, leading to superconductivity with $T_c=13.2$ K. Fluorination further suppresses the soft-phonon anomaly, yielding a moderate-coupling superconducting state with $T_c=5.27$ K. In contrast, W$_2$NCl$_2$ lies close to the CDW instability regime, exhibiting pronounced phonon softening that can be continuously tuned and ultimately suppressed by external strain or carrier doping. Under $-3\%$ compressive strain, the EPC constant is reduced to $\lambda=0.71$, resulting in a superconducting transition temperature of $T_c=5.8$ K.

Taken together, our results establish a unified framework in which CDW order and superconductivity emerge from the same soft-phonon-driven EPC mechanism. The balance between lattice stability and EPC strength determines whether the system favors a CDW phase, a superconducting state, or a regime where both tendencies coexist in close proximity. These findings identify W$_2$N-based monolayers as a versatile platform for engineering and controlling collective quantum phases in two-dimensional materials, and provide general insight into the relationship between soft phonons, CDW instabilities, and superconductivity in low-dimensional systems.

% =========================
% Acknowledgments
% =========================
\section*{Acknowledgments}
	This work was supported by the Second Century Fund (C2F), Chulalongkorn University (Grant No. C2F PD-2320260067). High-performance computing facility in this Research is funded by Thailand Science research and Innovation Fund Chulalongkorn University (ST-690022300001).

\bibliography{references}% Produces the bibliography via BibTeX.

@article{seeyangnok2026tunable,
  title={Tunable Superconductivity in Functionalized Janus MoSeA (A= H, Li) Monolayers: Competition between Lattice Instability and Electron Pairing},
  author={Seeyangnok, Jakkapat and Pinsook, Udomsilp and Ackland, Graeme J},
  journal={ACS Applied Energy Materials},
  volume={9},
  number={7},
  pages={4511--4522},
  year={2026},
  publisher={ACS Publications}
}

@article{jiang2023possible,
  title={Possible enhancement of the superconducting Tc due to sharp Kohn-like soft phonon anomalies},
  author={Jiang, Cunyuan and Beneduce, Enrico and Baggioli, Matteo and Setty, Chandan and Zaccone, Alessio},
  journal={Journal of Physics: Condensed Matter},
  volume={35},
  number={16},
  pages={164003},
  year={2023},
  publisher={IOP Publishing}
}

@article{seeyangnok2026moxhcdw,
  title={Charge Density Wave Order and Superconductivity in Janus MoXH Monolayers},
  author={Seeyangnok, Jakkapat and Pinsook, Udomsilp and Ackland, Graeme J},
  journal={arXiv preprint arXiv:2601.02959},
  year={2026}
}

@article{seeyangnok2024superconductivity,
  title={Superconductivity and electron self-energy in tungsten-sulfur-hydride monolayer},
  author={Seeyangnok, Jakkapat and Ul Hassan, M Munib and Pinsook, Udomsilp and Ackland, Graeme},
  journal={2D Materials},
  volume={11},
  number={2},
  pages={025020},
  year={2024},
  publisher={IOP Publishing}
}

@article{xue2024realization,
  title={Realization of hydrogenation-induced superconductivity in two-dimensional Ti 2 N MXene},
  author={Xue, Yamin and Cheng, Zebang and Yao, Shunwei and Wang, Ben and Jiang, Jiajun and Peng, Lin and Shi, Tingting and Chen, Jing and Liu, Xiaolin and Lin, Jia},
  journal={Physical Chemistry Chemical Physics},
  volume={26},
  number={35},
  pages={23240--23249},
  year={2024},
  publisher={Royal Society of Chemistry}
}

@article{lu2017janus,
  title={Janus monolayers of transition metal dichalcogenides},
  author={Lu, Ang-Yu and Zhu, Hanyu and Xiao, Jun and Chuu, Chih-Piao and Han, Yimo and Chiu, Ming-Hui and Cheng, Chia-Chin and Yang, Chih-Wen and Wei, Kung-Hwa and Yang, Yiming and others},
  journal={Nature nanotechnology},
  volume={12},
  number={8},
  pages={744--749},
  year={2017},
  publisher={Nature Publishing Group UK London}
}

@article{methfessel1989high,
  title={High-precision sampling for Brillouin-zone integration in metals},
  author={Methfessel, MPAT and Paxton, AT},
  journal={physical review B},
  volume={40},
  number={6},
  pages={3616},
  year={1989},
  publisher={APS}
}

@article{seeyangnok2024superconductivitywseh,
  title={Superconductivity and strain-enhanced phase stability of Janus tungsten chalcogenide hydride monolayers},
  author={Seeyangnok, Jakkapat and Pinsook, Udomsilp and Ackland, Graeme J},
  journal={Physical Review B},
  volume={110},
  number={19},
  pages={195408},
  year={2024},
  publisher={APS}
}

@article{pinsook2024analytic,
  title={Analytic solutions of Eliashberg gap equations at superconducting critical temperature},
  author={Pinsook, Udomsilp and Natkunlaphat, Nattawut and Rientong, Komkrit and Tasee, Pakin and Seeyangnok, Jakkapat},
  journal={Physica Scripta},
  volume={99},
  number={6},
  pages={065211},
  year={2024},
  publisher={IOP Publishing}
}

@article{liu2022two,
  title={Two-gap superconductivity in a Janus MoSH monolayer},
  author={Liu, Peng-Fei and Zheng, Feipeng and Li, Jingyu and Si, Jian-Guo and Wei, Liuming and Zhang, Junrong and Wang, Bao-Tian},
  journal={Physical Review B},
  volume={105},
  number={24},
  pages={245420},
  year={2022},
  publisher={APS}
}

@article{giannozzi2009quantum,
  title={QUANTUM ESPRESSO: a modular and open-source software project for quantum simulations of materials},
  author={Giannozzi, Paolo and Baroni, Stefano and Bonini, Nicola and Calandra, Matteo and Car, Roberto and Cavazzoni, Carlo and Ceresoli, Davide and Chiarotti, Guido L and Cococcioni, Matteo and Dabo, Ismaila and others},
  journal={Journal of physics: Condensed matter},
  volume={21},
  number={39},
  pages={395502},
  year={2009},
  publisher={IOP Publishing}
}

@article{schlipf2015optimization,
  title={Optimization algorithm for the generation of ONCV pseudopotentials},
  author={Schlipf, Martin and Gygi, Fran{\c{c}}ois},
  journal={Computer Physics Communications},
  volume={196},
  pages={36--44},
  year={2015},
  publisher={Elsevier}
}

@article{monkhorst1976special,
  title={Special points for Brillouin-zone integrations},
  author={Monkhorst, Hendrik J and Pack, James D},
  journal={Physical review B},
  volume={13},
  number={12},
  pages={5188},
  year={1976},
  publisher={APS}
}

@article{allen1975transition,
  title={Transition temperature of strong-coupled superconductors reanalyzed},
  author={Allen, Ph B and Dynes, RC},
  journal={Physical Review B},
  volume={12},
  number={3},
  pages={905},
  year={1975},
  publisher={APS}
}

@article{bekaert2019hydrogen,
  title={Hydrogen-induced high-temperature superconductivity in two-dimensional materials: The example of hydrogenated monolayer MgB 2},
  author={Bekaert, Jonas and Petrov, Mikhail and Aperis, Alex and Oppeneer, Peter M and Milo{\v{s}}evi{\'c}, MV},
  journal={Physical Review Letters},
  volume={123},
  number={7},
  pages={077001},
  year={2019},
  publisher={APS}
}

@book{peierls1955quantum,
  title={Quantum theory of solids},
  author={Peierls, Rudolf Ernst},
  year={1955},
  publisher={Oxford University Press}
}

@book{gruner2018density,
  title={Density waves in solids},
  author={Gruner, George},
  year={2018},
  publisher={CRC press}
}

@article{johannes2008fermi,
  title={Fermi surface nesting and the origin of charge density waves in metals},
  author={Johannes, MD and Mazin, II},
  journal={Physical Review B—Condensed Matter and Materials Physics},
  volume={77},
  number={16},
  pages={165135},
  year={2008},
  publisher={APS}
}

@article{calandra2011charge,
  title={Charge-density wave and superconducting dome in TiSe 2 from electron-phonon interaction},
  author={Calandra, Matteo and Mauri, Francesco},
  journal={Physical review letters},
  volume={106},
  number={19},
  pages={196406},
  year={2011},
  publisher={APS}
}

@article{weber2011extended,
  title={Extended phonon collapse and the origin of the charge-density wave in 2 H-NbSe 2},
  author={Weber, F and Rosenkranz, S and Castellan, J-P and Osborn, R and Hott, R and Heid, R and Bohnen, K-P and Egami, T and Said, AH and Reznik, D},
  journal={Physical review letters},
  volume={107},
  number={10},
  pages={107403},
  year={2011},
  publisher={APS}
}

@article{xi2015strongly,
  title={Strongly enhanced charge-density-wave order in monolayer NbSe2},
  author={Xi, Xiaoxiang and Zhao, Liang and Wang, Zefang and Berger, Helmuth and Forr{\'o}, L{\'a}szl{\'o} and Shan, Jie and Mak, Kin Fai},
  journal={Nature nanotechnology},
  volume={10},
  number={9},
  pages={765--769},
  year={2015},
  publisher={Nature Publishing Group UK London}
}

@article{ugeda2016characterization,
  title={Characterization of collective ground states in single-layer NbSe 2},
  author={Ugeda, Miguel M and Bradley, Aaron J and Zhang, Yi and Onishi, Seita and Chen, Yi and Ruan, Wei and Ojeda-Aristizabal, Claudia and Ryu, Hyejin and Edmonds, Mark T and Tsai, Hsin-Zon and others},
  journal={Nature Physics},
  volume={12},
  number={1},
  pages={92--97},
  year={2016},
  publisher={Nature Publishing Group UK London}
}

@article{zhu2017misconceptions,
  title={Misconceptions associated with the origin of charge density waves},
  author={Zhu, Xuetao and Guo, Jiandong and Zhang, Jiandi and Plummer, EW},
  journal={Advances in Physics: X},
  volume={2},
  number={3},
  pages={622--640},
  year={2017},
  publisher={Taylor \& Francis}
}

@article{sui2025two,
  title={Two-dimensional Janus MoSeH with tunable charge density wave, superconductivity and topological properties},
  author={Sui, Chang-Hao and Qiao, Shu-Xiang and Ding, Hao and Jiang, Kai-Yue and Shang, Shu-Ying and Lu, Hong-Yan},
  journal={Materials Today Physics},
  volume={53},
  pages={101698},
  year={2025},
  publisher={Elsevier}
}

@article{qiao2024prediction,
  title={Prediction of charge density wave, superconductivity and topology properties in two-dimensional Janus 2H/1T-WXH (X= S, Se)},
  author={Qiao, Shu-Xiang and Jiang, Kai-Yue and Sui, Chang-Hao and Xiao, Peng-Cheng and Jiao, Na and Lu, Hong-Yan and Zhang, Ping},
  journal={Materials Today Physics},
  volume={46},
  pages={101485},
  year={2024},
  publisher={Elsevier}
}

@article{han2023high,
  title={High-temperature superconductivity in two-dimensional hydrogenated titanium diboride: Ti2B2H4},
  author={Han, Yu-Lin and others},
  journal={Materials Today Physics},
  volume={30},
  pages={100954},
  year={2023},
  publisher={Elsevier}
}

@article{liu2024three,
  title={Three-gap superconductivity with T c above 80 K in hydrogenated 2D monolayer LiBC},
  author={Liu, Hao-Dong and Wang, Bao-Tian and Fu, Zhen-Guo and Lu, Hong-Yan and Zhang, Ping},
  journal={Physical Review Research},
  volume={6},
  number={3},
  pages={033241},
  year={2024},
  publisher={APS}
}

@article{ku2023ab,
  title={Ab initio investigation of charge density wave and superconductivity in two-dimensional Janus 2 H/1 T-MoSH monolayers},
  author={Ku, Ruiqi and Yan, Luo and Si, Jian-Guo and Zhu, Songyuan and Wang, Bao-Tian and Wei, Yadong and Pang, Kaijuan and Li, Weiqi and Zhou, Liujiang},
  journal={Physical Review B},
  volume={107},
  number={6},
  pages={064508},
  year={2023},
  publisher={APS}
}

@article{perdew1996generalized,
  title={Generalized gradient approximation made simple},
  author={Perdew, John P and Burke, Kieron and Ernzerhof, Matthias},
  journal={Physical review letters},
  volume={77},
  number={18},
  pages={3865},
  year={1996},
  publisher={APS}
}

@article{li2024machine,
  title={Machine learning accelerated discovery of superconducting two-dimensional Janus transition metal sulfhydrates},
  author={Li, Jingyu and Wei, Liuming and Shi, Xianbiao and Shi, Lanting and Si, Jianguo and Liu, Peng-Fei and Wang, Bao-Tian},
  journal={Physical Review B},
  volume={109},
  number={17},
  pages={174516},
  year={2024},
  publisher={APS}
}

@article{seeyangnok2025competition,
  title={Competition between superconductivity and ferromagnetism in 2D Janus MXH (M= Ti, Zr, Hf, X= S, Se, Te) monolayer},
  author={Seeyangnok, Jakkapat and Pinsook, Udomsilp and Ackland, Graeme J},
  journal={Journal of Alloys and Compounds},
  volume={1033},  
  pages={180900},
  year={2025},
  publisher={Elsevier}
}

@article{baroni2001phonons,
  title={Phonons and related crystal properties from density-functional perturbation theory},
  author={Baroni, Stefano and De Gironcoli, Stefano and Dal Corso, Andrea and Giannozzi, Paolo},
  journal={Reviews of modern Physics},
  volume={73},
  number={2},
  pages={515},
  year={2001},
  publisher={APS}
}

@article{balandin2021charge,
  title={Charge-density-wave quantum materials and devices—New developments and future prospects},
  author={Balandin, Alexander A and Zaitsev-Zotov, Sergei V and Gr{\"u}ner, George},
  journal={Applied Physics Letters},
  volume={119},
  number={17},
  year={2021},
  publisher={AIP Publishing}
}

@article{hwang2024charge,
  title={Charge density waves in two-dimensional transition metal dichalcogenides},
  author={Hwang, Jinwoong and Ruan, Wei and Chen, Yi and Tang, Shujie and Crommie, Michael F and Shen, Zhi-Xun and Mo, Sung-Kwan},
  journal={Reports on Progress in Physics},
  volume={87},
  number={4},
  pages={044502},
  year={2024},
  publisher={IOP Publishing}
}

@article{lian2023interplay,
  title={Interplay of charge ordering and superconductivity in two-dimensional 2 H group V transition-metal dichalcogenides},
  author={Lian, Chao-Sheng},
  journal={Physical Review B},
  volume={107},
  number={4},
  pages={045431},
  year={2023},
  publisher={APS}
}

@article{ali2025interplay,
  title={Interplay between charge-density-waves, disorder, and superconductivity in transition metal dichalcogenides},
  author={Ali, Aya and Alshaali, Fatima and Almheiri, Mejd and Abdel-Hafiez, Mahmoud},
  journal={Results in Physics},
  volume={70},
  pages={108154},
  year={2025},
  publisher={Elsevier}
}

@article{wang2023hydrogenation,
  title={Hydrogenation induced high-temperature superconductivity in two-dimensional W 2 C 3},
  author={Wang, Hao and Yin, Xin-Zhu and Liu, Yang and Li, Ya-Ping and Ni, Mei-Yan and Jiao, Na and Lu, Hong-Yan and Zhang, Ping},
  journal={Physical Chemistry Chemical Physics},
  volume={25},
  number={33},
  pages={22171--22178},
  year={2023},
  publisher={Royal Society of Chemistry}
}

@article{seeyangnok2025high,
  title={High-T c 2D ambient BCS superconductors in hydrogenated transition-metal borides},
  author={Seeyangnok, Jakkapat and Pinsook, Udomsilp and Ackland, Graeme J},
  journal={npj 2D Materials and Applications},
  volume={9},
  number={1},
  pages={70},
  year={2025},
  publisher={Nature Publishing Group UK London}
}

@article{seeyangnok2025ab,
  title={Ab initio investigation on structural stability and phonon-mediated superconductivity in 2D-hydrogenated M2X (M= Mo, V, Zr; X= C, N) MXene monolayer},
  author={Seeyangnok, Jakkapat and Pinsook, Udomsilp},
  journal={Journal of Physics and Chemistry of Solids},
  pages={113346},
  year={2025},
  publisher={Elsevier}
}

@article{meng2025mbenes,
  title={MBenes: A two-dimensional platform for high-temperature superconductivity via hydrogenation-induced Van Hove singularities},
  author={Meng, Xianghui and Shen, Yanqing and Yang, Xin and Cai, Kunpeng and Ai, Qing and Shuai, Yong and Zhou, ZhongXiang},
  journal={Physical Review B},
  volume={112},
  number={10},
  pages={104503},
  year={2025},
  publisher={APS}
}

@article{seeyangnok2026enhanced,
  title={Enhanced and Tunable Superconductivity Enabled by Mechanically Stable Halogen-Functionalized Mo2C MXenes},
  author={Seeyangnok, Jakkapat and Pinsook, Udomsilp},
  journal={arXiv preprint arXiv:2602.11552},
  year={2026}
}

@article{seeyangnok2026stability,
  title={Stability, Electronic Disruption, and Anisotropic Superconductivity of Hydrogenated Trilayer Metal Tetraborides (MB4H; M= Be, Mg, Ca, Al)},
  author={Seeyangnok, Jakkapat and Ackland, Graeme J and Pinsook, Udomsilp},
  journal={Advanced Theory and Simulations},
  volume={9},
  number={3},
  pages={e70356},
  year={2026},
  publisher={Wiley Online Library}
}

@article{seeyangnok2025phase,
  title={Phase stability and superconductivity in hydrogenated and lithiated Janus GaXS2 (X= Ga, In) monolayers},
  author={Seeyangnok, Jakkapat and Pinsook, Udomsilp},
  journal={Journal of Applied Physics},
  volume={138},
  number={16},
  year={2025},
  publisher={AIP Publishing}
}

@article{seeyangnok2025hydrogenation,
  title={Hydrogenation effects on the structural stability and superconducting properties of calcium-intercalated bilayer graphene C 2 CaC 2},
  author={Seeyangnok, Jakkapat and Pinsook, Udomsilp},
  journal={Nanoscale},
  volume={17},
  number={32},
  pages={18796--18804},
  year={2025},
  publisher={Royal Society of Chemistry}
}

@article{bhoi2016interplay,
  title={Interplay of charge density wave and multiband superconductivity in 2 H-Pd x TaSe2},
  author={Bhoi, D and Khim, S and Nam, W and Lee, BS and Kim, Chanhee and Jeon, B-G and Min, BH and Park, S and Kim, Kee Hoon},
  journal={Scientific reports},
  volume={6},
  number={1},
  pages={24068},
  year={2016},
  publisher={Nature Publishing Group UK London}
}

@article{seeyangnok2026competition,
  title={Competition between Charge Density Wave and Superconductivity in a Janus MXene Mo2NF2},
  author={Seeyangnok, Jakkapat and Pinsook, Udomsilp and Ackland, Graeme J},
  journal={arXiv preprint arXiv:2603.06284},
  year={2026}
}

@article{seeyangnok2026theoretical,
  title={Theoretical prediction of structural stability and superconductivity in Janus Ti 2 CSH MXene},
  author={Seeyangnok, Jakkapat and Pinsook, Udomsilp},
  journal={Physical Review B},
  volume={113},
  number={5},
  pages={054510},
  year={2026},
  publisher={APS}
}

@article{wang2023interplay,
  title={Interplay of the charge density wave transition with topological and superconducting properties},
  author={Wang, Zishen and You, Jing-Yang and Chen, Chuan and Mo, Jinchao and He, Jingyu and Zhang, Lishu and Zhou, Jun and Loh, Kian Ping and Feng, Yuan Ping},
  journal={Nanoscale Horizons},
  volume={8},
  number={10},
  pages={1395--1402},
  year={2023},
  publisher={Royal Society of Chemistry}
}

\end{document}